\documentclass[preprint,amsmath,amssymb,superscriptaddress]{revtex4-2}
\usepackage{graphicx}
\usepackage{dcolumn}
\usepackage{bm}
\usepackage{tikz}
\usetikzlibrary{quantikz}
\usepackage{caption}

\usepackage{booktabs}
\usepackage{hyperref}
\usepackage[mathlines]{lineno}%

\usepackage[normalem]{ulem} 
\usepackage{xcolor}          

\begin{document}

\title{Offline recovery of magic and entanglement from noisy Pauli product states}
\author{Erika Lloyd}
\email{erika.lloyd@sandboxquantum.com}
\affiliation{SandboxAQ}
\author{Alexandre Fleury}
 \affiliation{SandboxAQ}
\author{Marc P. Coons}
 \affiliation{Dow Core R\&D, Chemical Science}
 \author{James Brown }
 \affiliation{qBraid}
 \email{The work of J.B. was performed before joining qBraid}
 \author{Maritza Hernandez}

\begin{abstract}
The dependence of quantum algorithms on state fidelity is difficult to characterize analytically and is best explored experimentally as hardware scales and noisy simulations become intractable. While low fidelity states are often disregarded, they may still retain valuable information, as long as their dominant eigenvector approximates the target state. Through classical purification, we demonstrate the ability to recover resources specific to quantum computing such as magic and entanglement from noisy states generated by Pauli product formulas, which are common subroutines of many quantum algorithms. Additionally we show that the fidelity of the purified state is dependent on both the magnitude and order in which magic and entanglement are generated, which can be used to inform the order of operators within an ansatz. Consistent across simulation and experiment on IonQ's Aria quantum device, correlations within a state are found to be much more robust to noise than magic, and we show the advantage of designing algorithms targeting these low error states. This study uses quantum informatic tools for analyzing and optimizing quantum algorithms in a noisy framework.
\end{abstract}

\maketitle
\newpage 
\section{\label{sec:intro}Introduction}
A useful quantum algorithm must be competitive with classical methods which accomplishes the same task in a resource regime that is physically feasible~\cite{gonthier2022measurements,cerezo2023does}. Uncovering this area of utility will be much easier when universal digital quantum computers exist at a scale where meaningful benchmarks can be carried out. In the interim, quantum algorithm development is tasked with identifying the first potential areas where impact may occur. While we don't yet know the role of future quantum computers, we have tools for formulating quantum algorithms with promising scaling, and can observe their common subroutines~\cite{babbush2023exponential,low2024quantum,leimkuhler2025exponential}. We also know the specific properties that are unique to quantum computing, such as entanglement -- the presence of nonclassical correlations, and magic -- the degree of nonstabilizerness of a state. Both of these resources must be present in controllable amounts to go beyond the reach of classical algorithms~\cite{bravyi2019simulation,niroula2024phase}. Finally, we know that noise is an unavoidable feature of controlling matter at the quantum scale, and currently limits our ability to use these devices~\cite{chen2023complexity}. Considering this reality, current hardware presents an opportunity to understand basic features of the subroutines they may one day execute, and can give us insights on the relationship between the quantum resources we are targeting, and the environment we are shielding them from.

A large focus in quantum algorithms is electronic structure calculations where we would like to calculate the ground state energy of a system~\cite{whitfield2011simulation,cao2019quantum}. Candidate systems are typically strongly correlated which are difficult to simulate using classical quantum chemistry methods, and which are believed to be addressable quantumly~\cite{lee2023evaluating}. Algorithmic developments in Hamiltonian simulation~\cite{mukhopadhyay2023synthesizing} and state preparation using variational quantum algorithms~\cite{tang2021qubit, ryabinkin2018qubit} give us an indication of what types of operations may be useful. Many approaches rely on Pauli product formulas which execute subsequent applications of $e^{-i\theta P}$ where $P$ may act on any subset of qubits, and generates magic and entanglement based on the values $\theta$. It is not yet well understood how strong correlation of fermionic systems translate into resources such as entanglement or magic, as it depends highly on how the system is defined (e.g. delocalized vs localized molecular orbitals)~\cite{aliverti2024can} and the algorithm used.

 Quantum information metrics can be used as diagnostic tools for understanding the attributes of quantum algorithms. For example, re-framing the quantum Fourier transform as a time evolving Hamiltonian shows that this critical subroutine has small entanglement~\cite{chen2023quantum}. Other studies show that product-formula approximations methods have accelerated performance if the quantum dynamics are highly entangled, which may be relevant to strongly correlated systems~\cite{zhao2024entanglement}. But entanglement isn't enough to separate what is simulable classically from what is not, in fact too much entanglement is not useful at all~\cite{gross2009most}. A useful quantum computation must also contain magic, which distinguishes it from stabilizer states that are efficient to simulate classically~\cite{gottesman1998heisenberg}. Phase transitions between magic and entanglement have been observed in~\cite{fux2023entanglement, bejan2024dynamical} identifying the regimes that are not efficiently reachable by classical approaches such as matrix product states. For quantum approaches, there are also computational separations between entanglement dominated states which are accessible, and magic dominated states which are intractable~\cite{gu2403magic}.  Most quantum states either contain too much entanglement, or too much magic to be computationally useful~\cite{liu2022many}.

Most promising applications for quantum computers require access to a non-trivial state for optimal performance, but not all states are affected by noise equally. These effects are often ignored in algorithm development as a parallel task for error correction, but the interplay of noise with quantum resources can have interesting consequences. Recent studies have shown that coherent errors can generate high magic states~\cite{turkeshi2024coherent}, and that stabilizer measurements can either remove, or concentrate magic depending on the coherent error rate~\cite{niroula2024phase}. New efficient experimental protocols for measuring stabilizer entropy~\cite{leone2022stabilizer} have allowed for magic to be efficiently quantified in experiment~\cite{oliviero2022measuring}. There are also many works, including recent experiments~\cite{google2023measurement} showing that higher entanglement states have high error rates, which would ultimately increase error correction overhead. Purification methods~\cite{huggins2021virtual, koczor2021exponential, o2023purification} are a promising avenue for recovering noisy states without error correction. A missing component is understanding how recoverable states are based on their quantum resources such as magic or entanglement. This understanding can be a practical guide for what types of states we optimize hardware to produce, and may be an opportunity for incorporating classical methods~\cite{torlai2018neural,wei2024neural,kokaew2024bootstrapping} to extract useful information from noisy states.

In this work, we study the effects of noise on the magic and entanglement generated by states made using Pauli product formulas in both simulation and in experiment on an ion trap quantum device. We then demonstrate how these findings can impact quantum algorithmic design for quantum chemistry. In particular, we show a connection between the noise floor of a state after purification, and the magic within that state. We also show that product formulas which create an approximately equivalent state have a noise floor based on when they generate entanglement and magic during its construction. 

\section{Results}
This work explores the quantum resources generated by circuits typically used in quantum chemistry for initial state preparation and Hamiltonian simulation. It also explores the relationship between these resources and noise, using both simulations and experiments conducted on an ion trap quantum computer. The sections are organized as followed: In Section~\ref{sec:metrics} we introduce the metrics we will use to characterize quantum states. Section ~\ref{sec: circ_prim_sim} will then apply these tools to characterize a widely used circuit primitive in quantum algorithm in noisy simulation and experiment. In Section ~\ref{sec:Nontrivial_states} we generate non-trivial quantum states using randomized product formulas to explore the range of entanglement and magic these states occupy. We then explore the effects of noise on the order in which magic and entanglement are generated given the same target state. Section~\ref{sec:experiments} we show two case studies for simulating chemical systems on IonQ's Aria quantum device.

\subsection{Metrics for characterizing quantum states}
\label{sec:metrics}
Not all quantum states are useful, nor are they affected by noise equally.  To systematically characterize quantum states, we define a set of key metrics that serve as diagnostic tools for analyzing circuits used in quantum algorithms. These metrics are chosen for their interpretability, experimental accessibility, and relevance to algorithm design. A summary is provided in Table~\ref{tab:metrics}.

\begin{table}[h!]
    \centering
    \begin{tabular}{|l|c|c|}
        \hline
        \textbf{Metric} & \textbf{Measure}\\ \hline
        Observables & $\langle \mathcal{O}\rangle=tr(\mathcal{O}\rho)$ \\ \hline
        Magic Witness
         & $W_2 = -\log_2\bigg(\frac{1}{2^n}\sum_{P\in\mathcal{P}_n}| tr(\rho P)|^{4}\bigg) -3S_2(\rho)$  \\ \hline
        Quantum Mutual Information (QMI)&
        \begin{tabular}{@{}c@{}}  $QMI(A_1:A_2: ... :A_n) =$\\ $\sum_{k_i} [ S(X_{k_1},X_{k_1}...X_{k_n})] - (n-1)S(A_1,A_2...A_n)$ \end{tabular} \\ \hline
        Coherent Mismatch & $c = 1 - |\langle \psi_{ideal}|\psi_{purified}\rangle|^2$   \\ \hline
        Purity & $tr (\rho^2$) \\ \hline
        
    \end{tabular}
    \caption{Summary of metrics used in this work}
    \label{tab:metrics}
\end{table}

The method through which we extract information about a quantum state produced on a quantum computer is through measuring observables $\mathcal{O}$. These can be expressed in terms of Pauli words $P$. For a system of \(n\) qubits, a Pauli word is expressed as \(P = P_1 \otimes P_2 \otimes \cdots \otimes P_n\), where \(P_i \in \{I, X, Y, Z\}\) and $\mathcal{P}_n$ is the set of all $4^n$ Pauli strings. 

\subsubsection{Correlations and entanglement}
Entangling operations between qubits are a critical component in the design of quantum algorithms. In a noisy setting, entanglement can be destroyed, and additional correlations may be introduced. To study both the intended entanglement and the noise induced correlations, we quantify the total correlation of the quantum state generated by the quantum circuit. Correlation refers to the shared information between subsystems, and in the classical setting is quantified using mutual information. This has been naturally extend to the quantum setting. For a bi-partite state $\rho_{AB}$ the quantum mutual information can be calculated as:
\begin{equation}
    I(\rho_{AB}) = S(\rho_A) + S(\rho_B) - S(\rho_{AB}).
\end{equation}
Where $S(\rho) = -tr(\rho\log_2\rho)$ is the Von Neumann entropy, and $A$, $B$ refer to the subsystems of $\mathcal{H_A}\otimes \mathcal{H_B}$. This metric captures classical correlations, and quantum correlations $Q(\rho)$ which includes entanglement $E(\rho)$ but also other correlations beyond entanglement such as quantum discord. For general states, the correlation hierarchy is : $I(\rho) \geq Q(\rho) \geq  E(\rho)$~\cite{aliverti2024can}.  

Entanglement is defined as the non-separability of a state, and for a general mixed state, there is no closed form solution. For a pure state, entanglement can be calculated using the subsystem Von Neumann entropy of a bi-partition:
\begin{equation}
    E(|\psi\rangle \langle\psi|) = S(\rho_A) = S(\rho_B) = \frac{I(|\psi\rangle \langle\psi|)}{2}
\end{equation}
This equality shows that for a pure state, the quantum mutual information is the sum of the entropy for each subsystem. When discussing correlations in the pure state setting, we may discuss qualitative trends for QMI and entanglement interchangeably.

In this work we are considering the subsystems to be each individual qubit, and are interested in the correlations of the entire quantum state. For this reason we use the multiparty quantum mutual information (QMI) originally formulated by Watanabe et al.~\cite{Watanabe} and reformulated in \cite{kumar2017multiparty} to take into account the common information shared among parties.  Given a density matrix $\rho$ its QMI can be calculated using the following:
\begin{equation}
    QMI(A_1:A_2: ... :A_n) =\sum_{k_i} [ S(X_{k_1},X_{k_1},...,X_{k_n})] - (n-1)S(A_1,A_2,...,A_n).
\end{equation}
Where ${k_i}$ is the sum over distinct combinations of subsystems without repetition. The subsystems $X_{k_i}$ correspond to each qubit. The benefits of this measure are that it is strictly positive, and can be interpreted in a similar fashion as classical mutual information, which is the amount of useful information in the system. Second, by considering multipartite quantum mutual information, we can ignore the choice of bi-partition.

\subsubsection{Witnesses of magic in mixed quantum states}
Beyond entanglement, useful quantum computation also requires magic, the measure of the nonstabilizerness of the quantum state. This resource is necessary for the quantum computation to be beyond the reach of polynomial time classical algorithms, and is also the most costly resource to protect using quantum error correction. There are many witnesses for magic, in this work we will use the witness from Haug and Tarabungu~\cite{haug2504efficient} derived
from stabilizer Renyi entropies~\cite{leone2022stabilizer}. This witness is efficient to measure, and specifically meant to be used for mixed states:
\begin{equation}
W_\alpha(\rho) = \frac{1}{1-\alpha}\log_2 A_\alpha(\rho) -\frac{1-2\alpha}{1-\alpha}S_2(\rho).
\end{equation}
Here, $S_2(\rho)=-\log_2tr(\rho^2)$ is the 2-Renyi entropy, and $A_\alpha(\rho) = 2^{-n}\sum_{P\in\mathcal{P}_n} | tr(\rho P)|^{2\alpha}$ is the $\alpha$-moment of the Pauli spectrum,  representing the probability of a state being represented by a given Pauli string $P$. Any $W_\alpha(\rho)$ with $\alpha >1/2$ is a genuine witness of entanglement, and in this work we take $\alpha=2$ resulting in the following measure:
\begin{equation}
W_2 = -\log_2\bigg(\frac{1}{2^n}\sum_{P\in\mathcal{P}_n}| tr(\rho P)|^{4}\bigg) -3S_2(\rho).
\label{eq:magic}
\end{equation}
Higher values of  $W_2$ indicate a greater departure from stabilizer states $\rho_C$, while $W_2 \leq 0 $ indicates the state is a mixture of stabilizer states. For pure states, $W_\alpha$ recovers the stablizer entropies $\alpha$-Renyi entropies.

\subsubsection{Noise sensitivity and recovery metrics}
To capture the effect of noise on a given state, we track two noise-related metrics. Purity, defined as $tr \rho^2$, measures the degree of mixedness of a quantum state. A pure state has a purity of 1, while mixed states have lower values. Considering the growing prevalence of purification based error mitigation strategies \cite{huggins2021virtual, koczor2021exponential, o2023purification}, we also look at the limits of purification to recover quantum resources.  
 \begin{equation}
\psi_{purified}= \frac{\rho_{noisy}^M}{tr(\rho_{noisy}^M)}
\end{equation}
Where M is the order of purification. The central assumption for purification based error mitigation protocols is that the dominant eigenvector of the mixed state still corresponds to the desired state \cite{koczor2021dominant}. Perturbations of quantum hardware are not believed to be completely orthogonal to the target state, but any errors reduce the effectiveness of purification. The last metric we track is called the coherent mismatch~\cite{koczor2021dominant}, and quantifies the remaining coherent error in the state even after purification, a close variant has been referred to as the ``noise floor''~\cite{huggins2021virtual}, which we will use interchangeably.
\begin{equation}
    c = 1 - |\langle \psi_{ideal}|\psi_{purified}\rangle|^2
\end{equation}
In this work, when reporting the coherent mismatch, we first purify the state to order $M=5$, resulting in pure states to a high approximation. We selected this $M=5$ because preliminary sweeps shown in Fig~\ref{fig:purification_order} in Methods demonstrate that higher orders yield negligible improvements in state purity.

\subsection{Characterizing circuit primitives}
\label{sec: circ_prim_sim}
Algorithmic developments in Hamiltonian simulation and variational quantum algorithms provide insights into the types of operations that might be useful in future applications. Many approaches rely on Pauli product formulas which execute subsequent applications of $e^{-i\theta P}$ where $P$ may act on any subset of qubits. These operators serve as fundamental circuit primitives, forming the building blocks of quantum algorithms. Assemblies of these circuit primitives into unitary coupled cluster (UCC) operators and their relevant ground state energy estimates have been proposed as an application specific benchmark for quantum hardware \cite{mccaskey2019quantum}.  While measuring energy is critical for evaluating quantum algorithm performance in quantum chemistry, it conflates hardware performance with representational choices and measurement protocols for the observable. For evaluating hardware, it is more relevant to evaluate the state produced by hardware more directly using metrics reflecting their sought after quantum resources, such as magic and entanglement. These metrics will ultimately be useful if we can use insights gained to inform quantum algorithm development and improve their performance.

\begin{figure*}[hbt!]\includegraphics[scale=0.75]{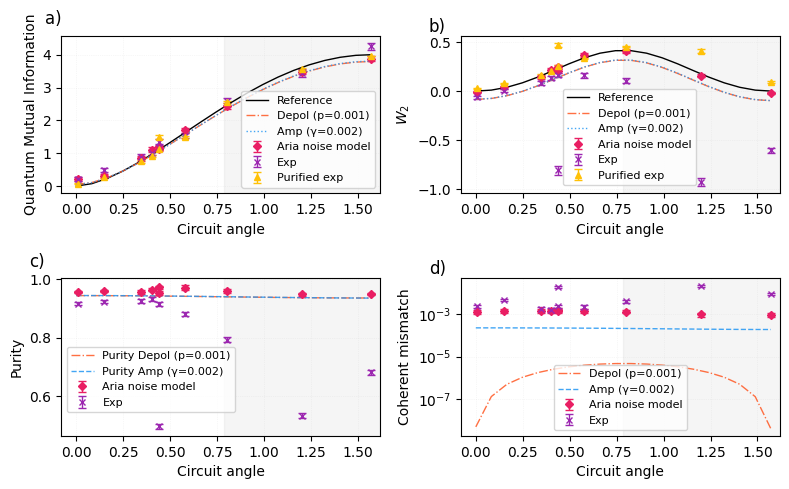}
    \caption{Simulated and experimental metrics of the circuit primitive measured as a function of circuit angle. Results are shown without noise (black line), with depolarization noise (dashed orange line), amplitude damping (dotted blue line), Aria noise model (red diamonds), experimental results (purple crosses), and purified experimental results (yellow triangles). The shaded area represents the region of "high entanglement" for a given value of magic. Metrics shown are a) the quantum mutual information b) the magic witness $W_2$ c) the state purity, and d) the coherent mismatch.}
    \label{fig:circuit_prim}
\end{figure*}

To illustrate how a circuit primitive generates quantum resources we consider $e^{-i\theta YXXX}$  applied to a reference state $\ket{1100}$ complied to the four qubit circuit shown Figure~\ref{cct: 4qubit} ~\cite{whitfield2011simulation}. The gates $H$ and $R_x$ correspond to rotations to and from the $Z$ basis to the basis of the qubit in $P$, and the single variational parameter $\theta$  in the $R_z$ gate determines the amount of magic and qubit correlation created by the circuit primitive. In Fig.~\ref{fig:circuit_prim} the black lines show a sweep of $\theta$ from $0 \rightarrow \pi/2$ and measures the QMI in a) and $W_2$ in b) simulated without noise. The amount of entanglement between qubits increases with circuit angle, and the amount of magic is maximal at $\theta = \pi/4$ and 0 at $\theta = 0, \pi/2$.  This means that for any angle, the circuit generates a unique amount of entanglement and magic. In Fig.~\ref{fig:circuit_prim} and subsequent studies on the circuit primitive, we shade the circuit angles above $\pi /4$ to indicate the region of ``high entanglement" for a given value of magic. 

\begin{figure}[h!]
\scalebox{0.8}{
\begin{quantikz}  
&&\lstick{\ket{1}} &\gate{R_x(\frac{\pi}{2})} & \ctrl{1}&\qw &\qw &\qw &\qw &\qw &  \ctrl{1} & \gate{R_x(\frac{-\pi}{2})} & \qw 
\\ &&\lstick{\ket{1}}& \gate{H}&  \targ{1} & \ctrl{1}&\qw &\qw &\qw &  \ctrl{1} & \targ{1} &\gate{H} & \qw 
\\ &&\lstick{\ket{0}} &\gate{H}& \qw & \targ{1}& \ctrl{1} & \qw & \ctrl{1} &  \targ{1} &\qw&\gate{H}& \qw 
\\ &&\lstick{\ket{0}}& \gate{H} &\qw & \qw & \targ{}&\gate{R_z(\theta)}  &\targ{} &\qw &\qw &\gate{H}& \qw   \end{quantikz}
}
  \caption{Quantum circuit implementation of $e^{-i\theta YXXX}$ applied to the initial state $|1100\rangle$.}
  \label{cct: 4qubit}
\end{figure}


To understand how noise impacts these resources, we show results obtained using a density matrix simulator with a depolarization channel (dashed orange line) and an amplitude damping channel (dotted blue line). In addition, we run simulations using a device noise model for IonQ's Aria~\cite{ionq_noise_2022} (red diamonds), and run experiments on the IonQ's Aria quantum computer~\cite{ionq2025aria} (purple crosses). For the device noise simulation and experimental data we approximate the density matrix $\rho$ using a single qubit Pauli classical shadow~\cite{huang2020predicting}, where experimental data was additionally postselected~\cite{jnane2024quantum}. Further details can be found in Methods~\ref{sec: exp_details}. Finally we also show these metrics after purifying the experimental state (yellow triangles). The purity of the noisy states are shown in Fig~\ref{fig:circuit_prim} c) and the coherent mismatch in d).

For our initial observations we choose a depolarization noise strength of $0.001$ and an amplitude damping of $0.002$ such that the purity of the simulated noisy states roughly aligns with the IonQ Aria noise model shown in Fig.~\ref{fig:circuit_prim} c). In Fig.~\ref{fig:circuit_prim} a) we observe that under noisy conditions QMI is generally well-recovered but is more susceptible to degradation in highly entangled states, with the exception of a few experimental points. The higher QMI at these points could be attributed to correlated errors not captured by the noise model.

In Fig.~\ref{fig:circuit_prim} b) the magic witness $W_2$ in the simulated results is more uniformly underestimated, regardless of the magnitude of magic in the states. For many points $W_2 < 0$, meaning it is indistinguishable from mixed stabilizer states. The experimental data deviates significantly more than predicted by any of the noisy simulations. We observe a clear correlation between the state purity and the magic witness, which is expected from Eq.~\ref{eq:magic} when the Reyni entropy is dominant. The experimental purity aligns well with the simulated device noise for low circuit angles, except for one point. For large circuit angles the purity of the state is significantly lower than any noise estimation. This discrepancy could stem from several factors: the timing of the experiment within the calibration window, increased correlated errors in states with higher entanglement, and the higher fidelity associated with small angle rotation gates~\cite{nam2020ground}. 

Interestingly the purified experimental values recover $W_2$ effectively, except for the points where it is overestimated. These points correspond to the highest coherent mismatch shown in ~\ref{fig:circuit_prim} d), suggesting the presence of additional coherent errors. In ~\ref{fig:circuit_prim} d) we see observe no strong relationship between coherent mismatch and circuit angle, except for the depolarization noise channel. In this case the coherent mismatch is much lower for a given state purity and increases with the amount of magic in the state.

\begin{figure*}[hbt!]\includegraphics[scale=0.55]{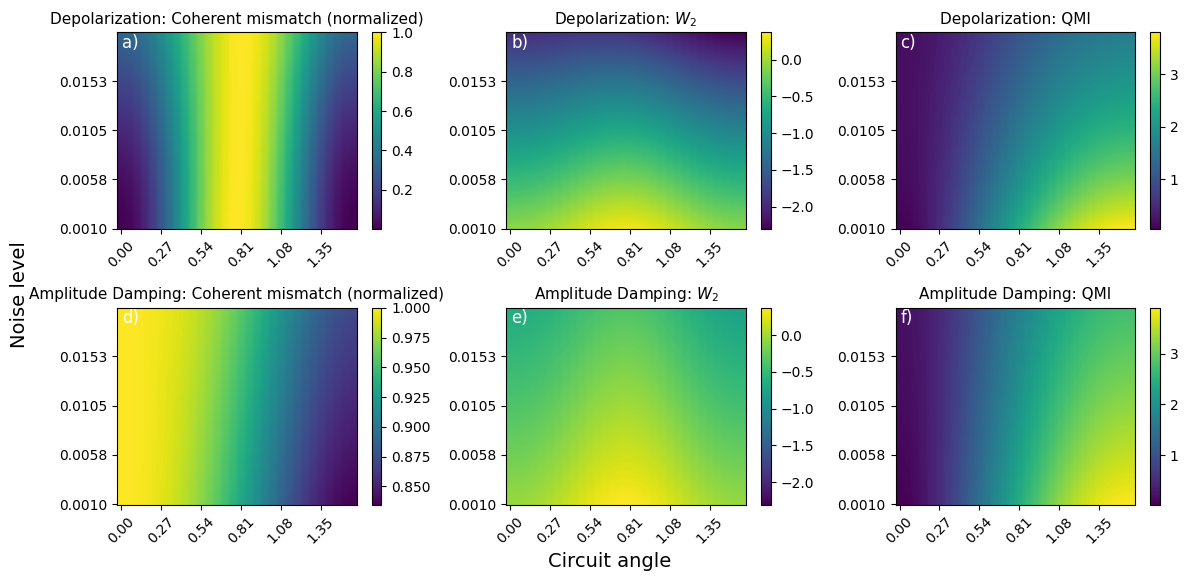}
    \caption{Simulations of circuit primitive depolarization (top row) and amplitude damping (bottom row) noise channels as a function of circuit angle for a range of noise strengths. The color bar represents the value of each if the following metrics: a) and d) coherent mismatch, b) and e) the magic witness $W_2$ c) and f) the quantum mutual information.}
    \label{fig:heatmaps}
\end{figure*}

To explore the distinct effects of depolarization and amplitude damping channels, in Fig.~\ref{fig:heatmaps} we plot the coherent mismatch normalized at each noise level, $W_2$, and QMI as function of circuit angle for increasing noise strengths. The relationship between the coherent mismatch and $W_2$ and QMI are different depending on the noise model. The depolarization model degrades magic more aggressively, and the coherent mismatch is correlated with the amount of magic in the target state. The coherent mismatch of the amplitude damping channel is instead anti-correlated with the QMI. These results show how quantum resources are effected differently based on noise channels, and show a difference in the ability for purification to recover states depending on noise type and resource. In the following section, we will explore whether these observed relationships hold for larger circuits constructed from these primitives. 

\subsection{Constructing quantum states using Pauli products}
\label{sec:Nontrivial_states}
Most promising applications for quantum computers require access to a non-trivial state for optimal performance. A non-trivial state should contain both magic, and entanglement in appropriate amounts such that it is difficult to simulate classically, and yet has useful properties. For a given unitary U, its construction is not unique and can be expressed in terms of $N$ smaller unitaries $g_i$.
\begin{equation}
    | \Psi \rangle = U|\psi_{ref}\rangle = g_N...g_2g_1|\psi_{ref}\rangle 
\end{equation}

To explore circuits relevant to quantum chemistry we will probe the entanglement and magic produced when each $g_i$ is a Pauli exponential $e^{-i\theta P}$, and $U$ is formed by Pauli products. We then look at a specific target state $| \Psi \rangle$ approximated with two different unitaries, and show that the order in which entanglement and magic are generated impact the noise floor of the generated state.

\subsubsection{Generating magic and entanglement with randomized Pauli product formulas}

\begin{figure*}[hbt!]   \includegraphics[scale=0.65]{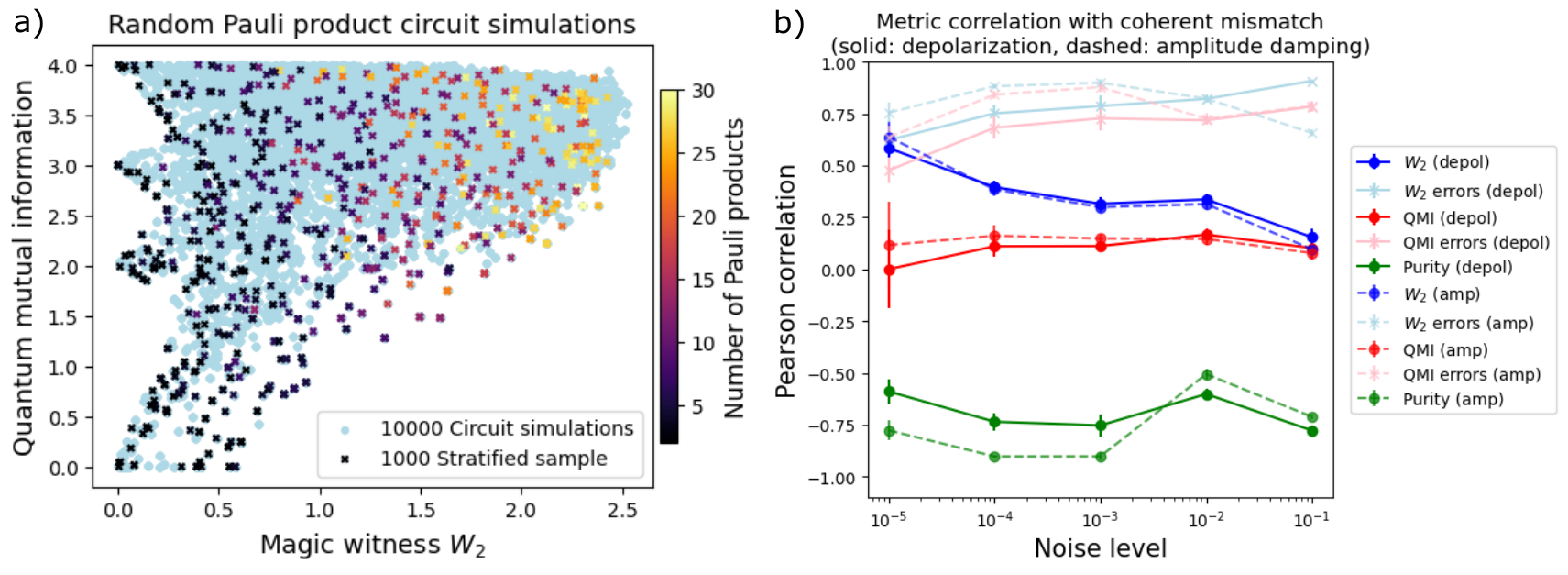}
    \caption{Random circuit simulations. We simulate 10000 random 4 qubit circuits by choosing between 2-30 random Pauli exponentials acting on either 2 or 4 qubits.  We track: the purity, the stabilizer entropy  ($W_2$) , and the multipartite quantum mutual information (QMI) of the ideal state and noisy states, and the coherent mismatch of the state after going through a depolarizing noise channel, and being purified to M=5. a) A scatter plot of the QMI and $W_2$ for all circuits, and a stratified sample. The color bar represents $N$, the number of Pauli products in the stratified sample. b) For each metric, Pearson correlations between the metric and the coherent mismatch are evaluated on each of 10 stratified subsets. The mean and standard deviation across subsets are plotted for a range of depolarization and amplitude damping noise strengths. $W_2$ errors and QMI errors are the absolute difference between the theoretical and noisy values of these metrics.}
    \label{fig:random_circuit}
\end{figure*}

In Fig.\ref{fig:random_circuit} we explore the magic and entanglement that may be generated from circuits made of Pauli product formulas. We run 10000 random circuit simulations on 4 qubits, each with the number of Pauli products $N$ uniformly sampled from 2 and 30, and which acts on a random uncorrelated reference state made from sampling $X$ or $H$ gates applied to the zero state. Each $P$ is a random Pauli word acting on either 2 or 4 qubits, and $\theta$ is uniformly sampled between $0$ and $2\pi$. Fig.\ref{fig:random_circuit} a) shows a scatter plot of the resulting QMI and $W_2$ for all 10000 circuit simulations, as well as a 1000 circuit subset using a stratified sampling method to mitigate undersampling biases in the simulations. The stratified subsets are made by making 10 bins for each of the QMI and $W_2$ data values, and then evenly sampling from each bin to reach 1000 circuits. The colorbar on the stratified sample represents $N$, the number of $g_i$ used to generate the corresponding state, and is a proxy for the relative depth of the circuits. We observe that the random Pauli products can easily generate states with high entanglement at low depth, whereas high magic states tend to be deeper. We are also unable to generate states with low entanglement and high magic.

We explore the connection of magic and entanglement within Pauli product states to the coherent mismatch in Fig.\ref{fig:random_circuit} b) by applying depolarizing (solid lines) and amplitude damping (dotted line) noise channels at increasing strengths to the 10000 random circuits. For 10 non-overlapping stratified subsets of size 1000, we calculate the mean and standard deviation of the Pearson correlation between the coherent mismatch and several metrics: the state purity, the theoretical values of QMI and $W_2$,  and the absolute errors in QMI and $W_2$ of the noisy states. As expected the purity is anti-correlated to coherent mismatch. The correlation of $W_2$ is high for low noise strength, and decreases as the noise strength increases. The QMI is nearly uncorrelated for all noise levels, but has high variance in the low noise region. This is likely due to low QMI being undersampled in this study. The absolute errors in QMI and $W_2$ are highly correlated to the coherent mismatch for all noise levels. This demonstrates the connection between the noise floor of the noisy quantum states, and their ability to accurately generate magic and entanglement.

This analysis suggests that as hardware improves, the noise floor of a quantum computation using Pauli product formulas will be dependent on the amount of magic we seek to produce, and more weakly related to the qubit correlations within the state. These results only look at the resources in the final state, but ignore intermediate steps. If quantum resources are easily corruptible by noise, this suggests the state may be more stable if magic and entanglement are generated near the end of the circuit.

\subsubsection{Operator ordering and the accumulation of noise}
Trotterization studies on the unitary coupled cluster ansatz~\cite{grimsley2019trotterized} demonstrate that the sequence in which unitaries are applied has a significant affect on energy estimations. In this study, we further show that unitary ordering also influences the accumulation of noise and performance of a circuit. To illustrate this,  we study two different approximations to the ground state of a stretched linear $H_3$ molecule, obtaining the reference $|\psi_{gstate}\rangle$ using exact diagonalization. To construct the first ground state circuit approximation, we use the involutory linear combinations  of anti-commuting Pauli generator (ILC) method developed in~\cite{ryabinkin2023efficient} and implemented in Tangelo~\cite{senicourt2022tangelo}. This method results in a $U_1$ consisting of 6 Pauli exponentials $g_i$. To find an alternative circuit $U_2$, we take the same set of $g_i$ and apply them in a different order. The ordering for this study was chosen because it diverged most strongly in its resource generation compared to the original unitary while maintaining reasonable Trotter error. In Methods~\ref{sec: unitary_paths_details} we show the set of $g_i$, and a set of alternative orderings this study could be extended to.  The overlaps of the states generated by $U_1$ and $U_2$ with the target state are 0.99997 and 0.99725, respectively. The corresponding errors in ground state energies are $1.85\times10^{-6}$ and $5\times10^{-4}$ Hartree, both well below chemical accuracy of $1.59$ mHa. We consider both $U_1$ and $U_2$ to be good approximations to the target state which use the same set of $g_i$, but with a different ordering.

\begin{figure*}[hbt!]   \includegraphics[scale=0.65]{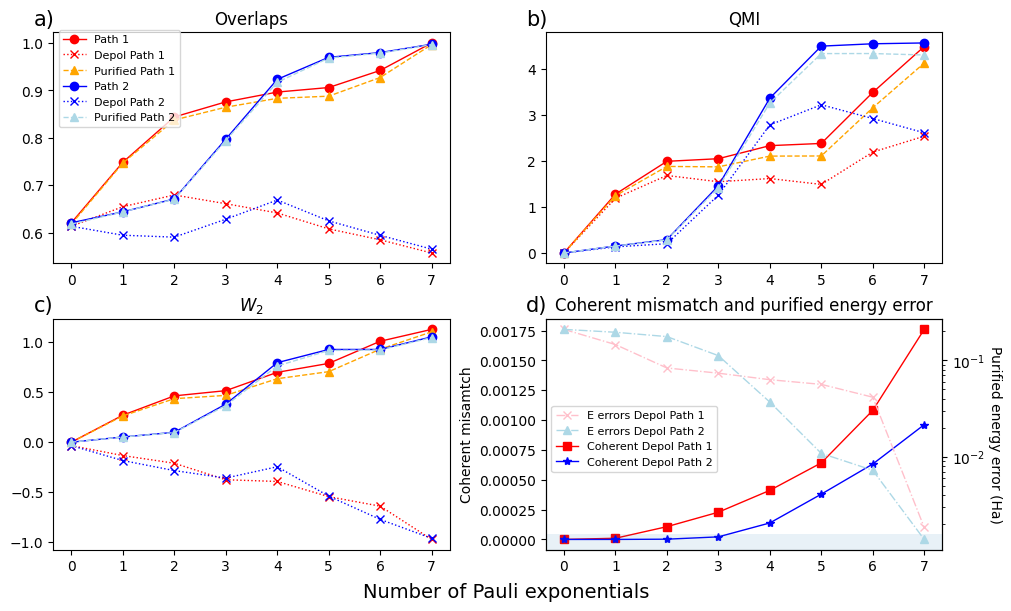}
    \caption{Metrics for two paths (red and blue) which generate approximations to the ground state of a linear $H_3$ molecule.  Metrics are measured after the application of each Pauli exponential $g_i$ where $n$ is marked on the x-axis. In figure a)-c), the solid lines represent the noiseless simulation, the dotted lines are simulations with a depolarization noise of 0.0005 applied after each Pauli exponential, and dashed lines are the values after purification.  a) Overlap with the target state. b) The quantum mutual information (QMI). c) The Magic witness ($W_2$) d) Coherent mismatch and the energy error of the purified state. The grey band represents values within a chemical accuracy of $1.59$ mHa. }
    \label{fig:circuit_path}
\end{figure*}

We study properties of the state produced after the application of each $g_i$ for both unitaries $U_i$, referring to each as a ``path" to the target state. In Fig.\ref{fig:circuit_path} the path 1 for $U_1$ is shown in red and path 2 for $U_2$ in blue. To mimic the accumulation of noise over time in a quantum experiment, a depolarization noise of strength 0.0005 is applied to the circuit after each $g_i$. In Fig.\ref{fig:circuit_path} a)-c), we plot the overlap with the target state, the quantum mutual information (QMI), and the stabilizer entropy  ($W_2$), where solid lines represent noiseless simulations, dotted lines indicate noisy simulations, and dashed lines are the purified values.  In Methods~\ref{sec: unitary_paths_details} we present additional studies exploring a range of noise strengths from 0.0005 to 0.001, as well as an amplitude damping channel, which exhibit similar trends. Here, for clarity of presentation, we include only the depolarization channel with a noise strength of 0.0005.

The overlaps start at the mean field reference $\langle\psi_{gstate}|\psi_{ref}\rangle$ and sequentially increases as each $g_i$ is applied. Due to the different ordering of $g_i$ in each unitary $U_i$, path 1 generates a large overlap with the target state in the first few $g_i$  0-2 applied Pauli exponentials. This is also reflected in the early generation of entanglement and magic. In the presence of noise the overlap plateaus around $g_3$ and slowly decreases. Path 2 generates most of its overlap with target state between $g_i$ 2-4, with its peak noisy overlap at $g_4$, after which it slowly decreases. 

Without any error mitigation, the best performing state in the noisy setting would be constructed by executing path 2 up to $g_4$ since it has the highest overlap with the target state (peak noisy overlap at $g_4$ of approximately 0.78 for Path 2 vs. 0.76 for Path 1). This is also the point with the largest $W_2$ in the noisy setting. We observe that despite the overlaps decreasing for the noisy states generated by the larger circuits, qubit correlations are still being generated in amounts qualitatively similar to the noiseless setting. Magic however is more difficult to generate throughout the circuit construction. To increase the magic in the state, each additional Pauli exponential $g_i$ must generate more magic than is lost due to the noise it introduces. This highlights the greater difficulty, and importance, of preserving magic in a state rather than the correlations. 

To explore the potential utility of these noisy low overlap states, we show in Fig~\ref{fig:circuit_path} d) the coherent mismatch of each path after purification, and the error in ground state energy calculated using the purified states. These purified states continue to approach the ground state after each $g_i$, and recover the ground state energy within chemical accuracy as indicated by the blue shaded region. This is also reflected in the purified values for $W_2$ and QMI which align well with the reference values. This demonstrates that the states are recoverable and can still be useful, given there is a scalable method for purifying them, or other method for extracting their properties. This has possible implications for quantum algorithms which have an initial state preparation component followed by quantum phase estimation (QPE), or other refinement method. These methods typically choose their initial state based on overlap alone, which may not be optimal in a noisy framework.

\begin{figure*}[hbt!]   \includegraphics[scale=0.5]{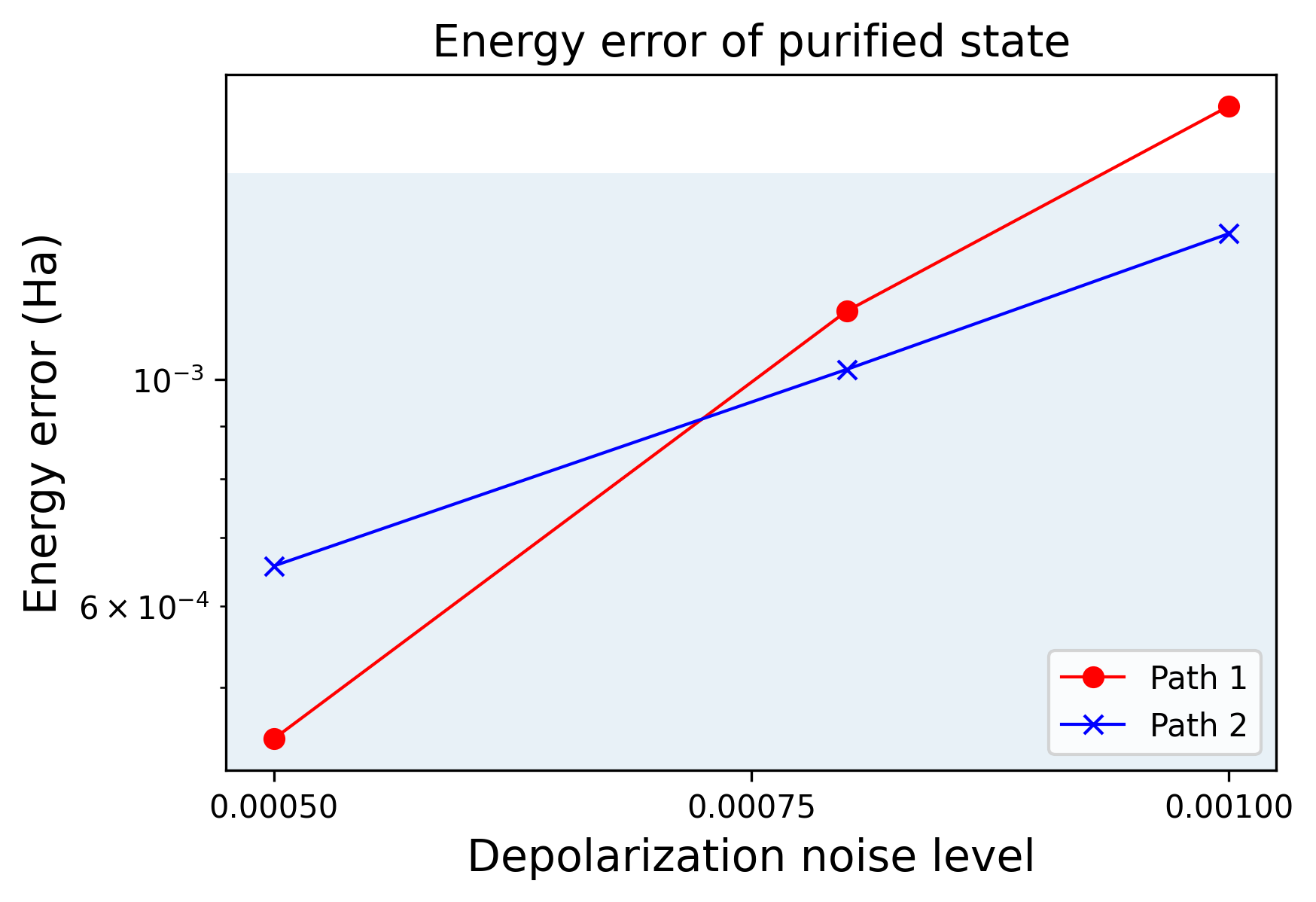}
    \caption{Energy errors of purified states generated by $U_1$ and $U_2$ for depolarization noise levels 0.0005, 0.0008 and 0.001 applied after each Pauli exponential. Blue shaded region indicates chemical accuracy for ground state energy of linear $H_3$ molecule.}
    \label{fig:energy_paths}
\end{figure*}

Interestingly, path 1 which generates magic and entanglement early in its circuit construction has a higher coherent mismatch throughout the entire path, even in the regions where path 2 has higher $W_2$ and QMI (at $g_i$ 4 and 5). This is also evident by looking at the greater errors in overlap, $W_2$, and QMI in the purified values after $g_2$ compared to path 2. This shows us that the order in which we generate resources has an impact on the noise floor of a quantum computation, and should be considered when compiling unitaries into gates. In Fig~\ref{fig:energy_paths} we demonstrate that for larger noise levels, the coherent mismatch can be significant enough such that path 2 results in more accurate energy estimates despite $U_2$ generating a worse approximation to the ground state in the noiseless setting. This study illustrates that a unitary with higher overlap to the target state does not necessarily lead to better performance if its resources are less protected from noise. Instead, a unitary with a lower noise floor can result in improved energy estimates and state overlaps. This sensitivity to operator ordering can be incorporated into algorithms with Trotterization protocols, and with dynamic ansätze~\cite{ryabinkin2020iterative, grimsley2019adaptive} by selecting operator sequences that defer resource generation to later stages of the circuit.

\subsection{Experimental case studies ion trap quantum computer}
\label{sec:experiments}
Studying how noise impacts entanglement and magic on hardware is important for understanding the feasibility of quantum algorithms on near-term devices. To extend the findings from our numerical simulations to an experimental setting, we perform experiments on IonQ's ion trap device, Aria~\cite{maksymov2023enhancing}. These experiments study how quantum resources degrade in the presence of real hardware noise, and the extent to which purification techniques can recover them. Additionally, we explore  how error rates obtained from postselection serve as an efficient proxy for evaluating which types of circuits are best suited for being executed on quantum hardware. As a practical demonstration, we show an example of tailoring a range of chemical systems to be computable by a single state identified to interact most favorably with hardware noise.

\subsubsection{Tailoring Hamiltonian to lowest error circuit primitive}
\begin{figure*}[hbt!]   \includegraphics[scale=0.6]{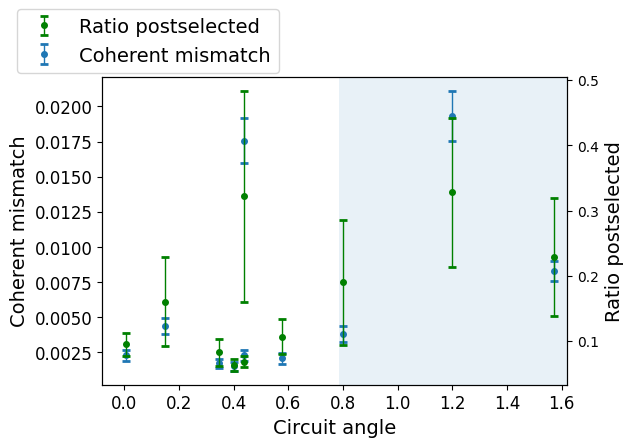}
    \caption{Coherent mismatch (blue) and ratio of postselected data (green) for 4 qubit experiment $|\psi\rangle=e^{-i\theta YXXX}|1100\rangle$ on IonQ's Aria device as a function of $\theta$. }
    \label{fig:errors}
\end{figure*}

We first analyze the experimental results obtained with the 4-qubit circuit primitive shown in Fig.~\ref{fig:circuit_prim}. To assess the practical utility of the resulting noisy states, we study the dissociation curve of $H_2$ in the STO-3G basis. A 4-qubit system is obtained using the Jordan Wigner fermion to qubit mapping, and is solvable using the ansatz $|\psi\rangle=e^{-i\theta YXXX}|1100\rangle$ with a single variational parameter $\theta$. In this case study we will demonstrate that dressing the Hamiltonian $H^*=D^\dag HD$ with an appropriate operator $D$ it can be made compatible with any circuit executed by the experiment $U_{circ}$. 
\begin{equation}
 E = \langle0| U^\dag_{ansatz}H U_{ansatz}|0\rangle = \langle 0 | U^\dag_{circ} D^\dag H D U_{circ}| 0 \rangle =  \langle 0 | U^\dag_{circ} H^* U_{circ}| 0 \rangle 
\label{eq:tailor}
\end{equation}
The effect of this $D$ is to shift the correlation between the circuit and Hamiltonian. Figure~\ref{fig:errors} shows the coherent mismatch (blue) and the ratio of postselected data (green) as functions of the circuit angle, extracted from the experimental measurements.

The following study shows that by choosing $U_{circ}$ to correspond with the lowest ratio of postselected data, we will yield more accurate results than using the circuit initially identified through the variational quantum eigensolver (VQE) optimization $U_{ansatz}$. Details of our method for identifying the dressing operator $D$ can be found Methods ~\ref{sec:tailoring_hamiltnoian}. We calculate the dissociation curve where the energy at each atomic distance $R$ is solved using a unique angle $\theta$. We then identified the circuit with the lowest ratio of postselected data to be $0.06$ when $\theta = 0.401$, which also corresponds to the circuit with the lowest error in magic. Using Eq.\ref{eq:tailor}, the Hamiltonians of each point along the curve are dressed with an operator $D$ to be computable by the state produced by $|\psi\rangle=e^{-i0.401 YXXX}|1100\rangle$. 
\begin{figure*}[h]   \includegraphics[scale=0.35
   ]{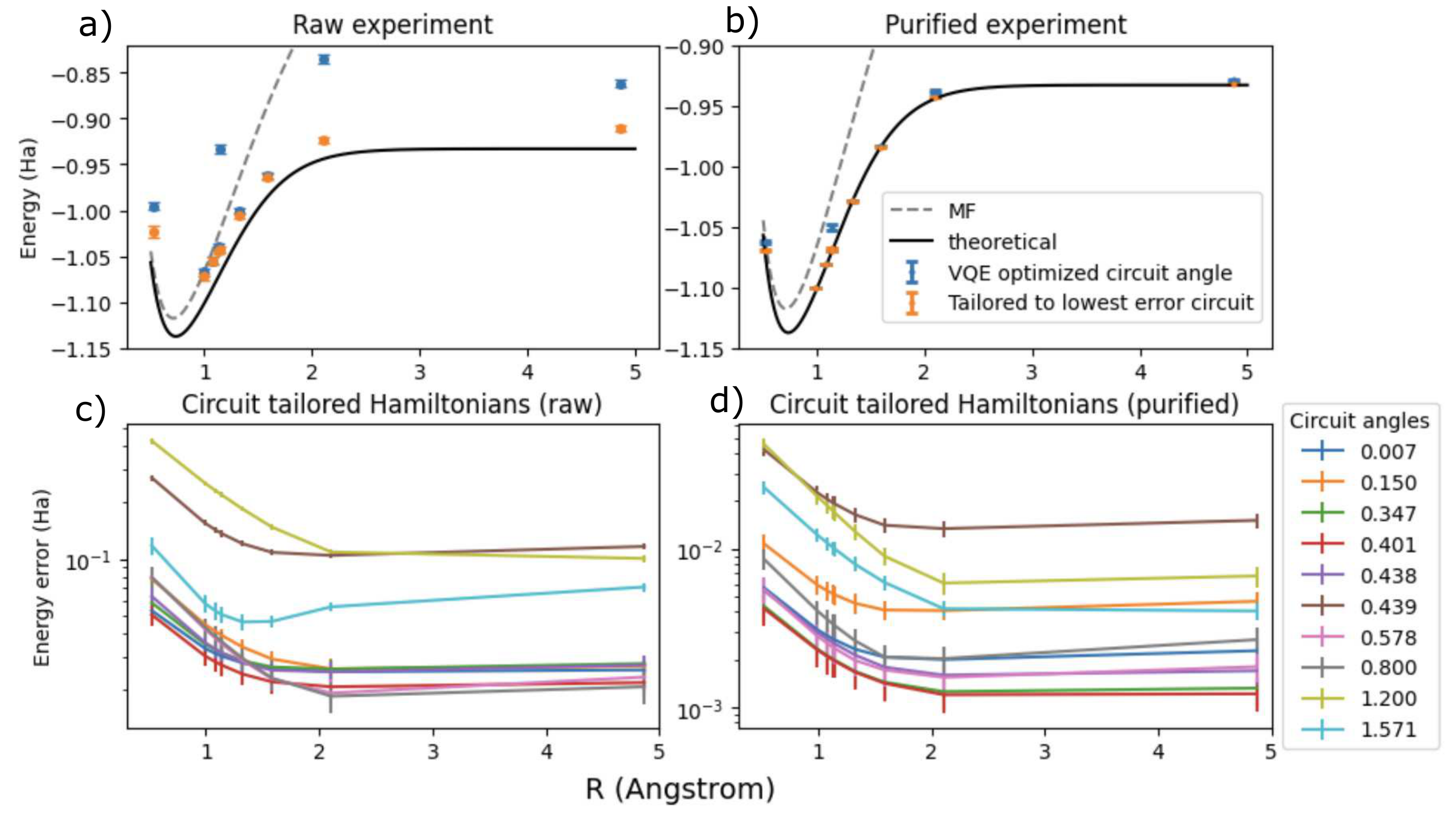}
    \caption{Experimental measurements of $H_2$ dissociation curve with circuit tailored Hamiltonians. Raw experimental (left column) and purified experimental (right column). Reference FCI, and mean field (MF) energies are shown with the black and black dotted line respectively. a-b) Experimental results with tailoring $0.0401$  (orange) and without (blue). c-d) Energy error along the dissociation curve if tailoring to each different circuit angle.}
     \label{fig:H2_tailored}
\end{figure*}

The dissociation curves in Fig.\ref{fig:H2_tailored} a) and b) show the results obtained from the raw and purified experimental data, respectively. In both cases, tailoring to the lowest-error circuit (orange) yields better agreement than the original VQE optimized circuit (blue). Although dressing the Hamiltonian increases the number of terms required for a single-point energy evaluation, it can be advantageous when performing many calculations, as the corresponding expectation values can be reused. In this example, using a separately optimized VQE circuit at each geometry along the dissociation curve would require measuring a total of 150 Hamiltonian terms. In contrast, with Hamiltonian dressing we employ a fixed circuit for the entire curve, so each point reuses the same set of expectation values. As a result, the total number of Hamiltonian terms that must be measured for the entire dissociation curve is reduced to 19.

In Fig.\ref{fig:H2_tailored} c) and d), we show how the energy error from the reference differs for each circuit in our experimental test set for the raw and purified data. Overall, the state at angle $0.401$ performs the best in both settings, with some better candidate circuits (0.578 and 0.8) in the raw case for the points at larger distance $R$.This shows the power of a single well chosen noisy state, and that we can have improved performance by tailoring our methods to be compatible with the states that are more stable in the presence of hardware noise.

\subsubsection{Circuit path for Be atom}
To examine the noise resilience of quantum resource generation for more complex states, we increase the depth of the circuit. In this next experiment, magic and entanglement is increased by adding Pauli exponentials, rather than sweeping the parameter angles in the circuit. The target is to approach the ground state of the beryllium (Be) atom after applying a problem decomposition approach based on the method of increments (See Methods~\ref{sec: MIFNO} for details). 

\begin{figure*}[hbt!]   \includegraphics[scale=0.5
   ]{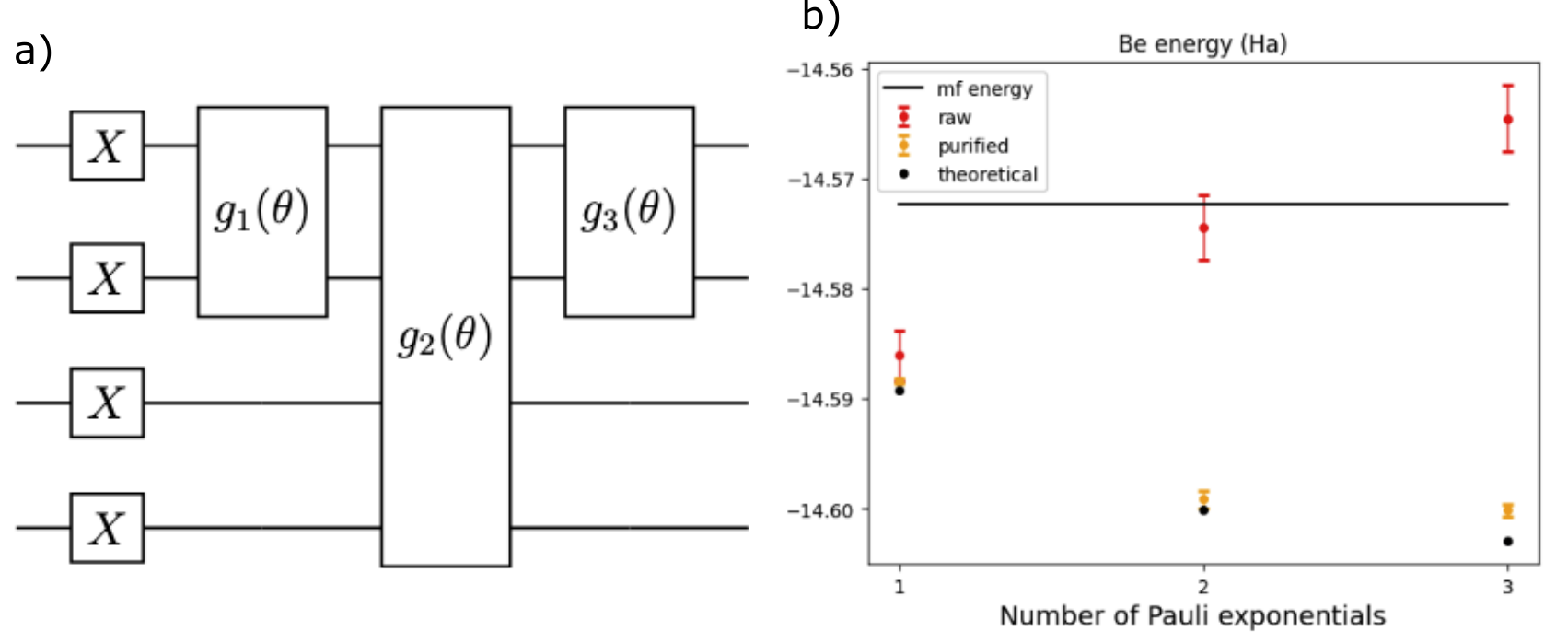}
    \caption{Experimental implementation of larger circuits targeting the ground state of the Be atom. a) Circuit construction using sequential unitary operations $g_1, g_2, g_3$.
    b) Raw (red) and purified (orange) energy estimates at different circuit depths.}
     \label{fig:Be_depth}
\end{figure*}

Similar to the simulations, we construct the unitary U from circuit primitives $g_1, g_2, g_3$ applied to a reference state as depicted in the circuit in Fig.\ref{fig:Be_depth} a), and further circuit details are in Methods~\ref{sec: exp_circuit_details}. We run three different experiments which characterize the state at each step after applying a unitary $g_i$ and applying a symmetry verification circuit. In Fig.\ref{fig:Be_depth} b) we present the raw (red) and purified (orange) values for the experimental energy. The raw data shows that as additional unitaries are applied, the measured energy deviates further from the expected ground state value, eventually surpassing the mean-field energy estimate. This trend is mitigated in the purified data, which follows the expected theoretical behavior. Similar to what was observed in the simulations, this dramatic change in energy estimation indicates that useful resources are likely still being generated in the raw state.

\begin{figure*}[hbt!]   \includegraphics[scale=0.65
   ]{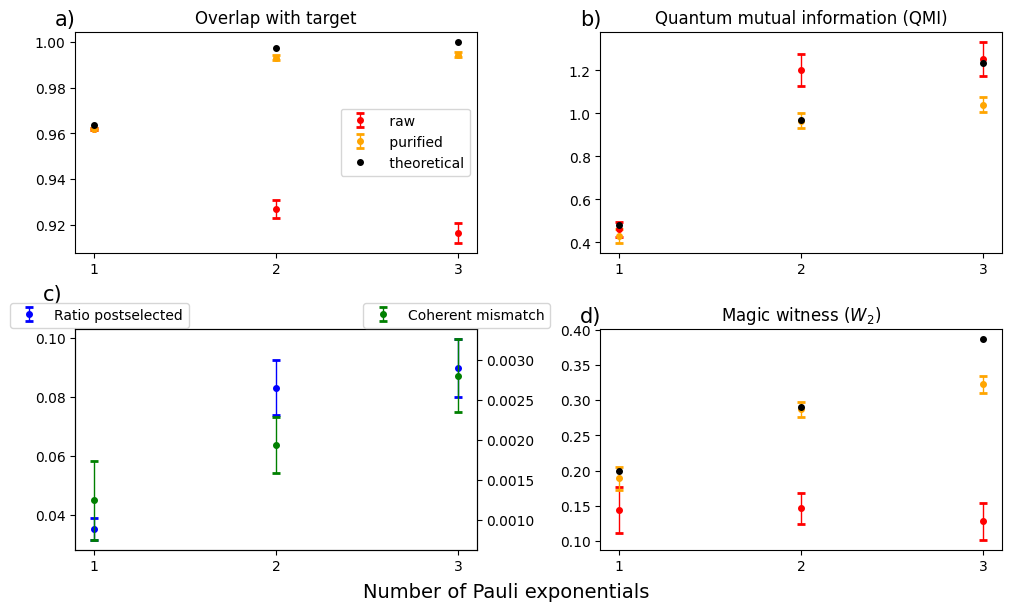}
    \caption{Experimental metrics for approaching the target state of the Be atom are shown as a function of the number of applied Pauli exponentials. Results are presented for the raw (red), purified (orange), and reference (black) states. (a) Overlap with target; (b) Quantum mutual information (QMI); (c) Coherent mismatch (green) along with the ratio of postselected data (blue); and (d) Magic witness ($W_2$);}
     \label{fig:exp_path_SE_QMI}
\end{figure*}

To explore further, we calculate several metrics shown in Fig~\ref{fig:exp_path_SE_QMI}. Raw (red), purified (orange), and theoretical (black) results for overlap with the target state are shown in a), QMI in b) and $W_2$ in d). Coherent mismatch is plotted in green, and ratio of postselected data in blue in d). We see that as we successively apply unitaries to our circuit in the noiseless setting we begin to approach the target state. In the raw experiment, each increase in depth of the circuit results in a decrease in overlap -- this is mirrored in the increase in the ratio of postselected data. Purifying the states has a large effect on the overlaps (orange), suggesting that despite the unitaries adding a significant amount of noise, there is still recoverable information from the state. We can also see that despite the coherent mismatch being the largest for the deepest circuit, it was not enough to offset the overlap gained by applying the final noisy unitary. This is also reflected in the purified energy in Fig~\ref{fig:Be_depth} b). If we were to continue approaching the target state, the overlap of the purified state would eventually be limited by the noise floor.

The significant effect of purification is also seen in the magic witness $W_2$ whose raw values are low and do not increase with depth of the circuit. Upon purification there is a dramatic improvement towards the reference values. The raw QMI values either align well with the expected values, or over estimate them.  The purified values lower the QMI in all cases, even to below expected values in the cases where it originally aligned. This suggests that the extra correlations contributing to the measured QMI may be classical, or from another resource such as quantum discord. In this experiment, the state correlations are more robust than the magic, whose degradation results in low overlap with the target state. Purification brings the state closer to the target, and this improvement is reflected in the recovery of magic. These observations reinforce the conclusions from our simulations: magic is the more challenging resource to preserve and has the greatest impact on energy calculations. Regardless of these differences, we are still able to recover entanglement and magic offline in a meaningful way without tailoring the measurement strategy to extract these quantities apriori. This has allowed us to analyze the quantum states in a more detailed way than energy estimates alone, and may enable the design of efficient classical or quantum methods which extract and utilize information from noisy quantum states. 

\section{Conclusions}
In this work we have explored the effects of noise on resources such as magic and entanglement of a quantum state. We demonstrate that traditional metrics such as overlap with target state, or ground state energy estimates, do not fully capture the characteristics of quantum states in a noisy setting. In particular, we see that entanglement is relatively robust compared to magic, whose degradation is reflected in the decrease in overlap. We also demonstrate that by purifying noisy quantum states we can recover both their magic and entanglement, enabling more accurate calculation of the corresponding energy estimates. While this offline purification is not scalable, we believe it to be a useful diagnostic tool for small circuits executable on current hardware for studying the effects of noise. In the future we are optimistic that more sophisticated postprocessing protocols may be able to extract properties of the noisy dominant eigenvector more directly and efficiently. Promising avenues include robust classical shadow methods~\cite{chen2021robust,jnane2024quantum} coupled with machine learning~\cite{huang2210learning, kokaew2024bootstrapping}. As hardware continues to mature, quantum distillation methods can be incorporated into the circuit, reducing the classical cost. When designing algorithms in the noisy context, we may consider evaluating their performance based on the properties of their dominant eigenvector, as opposed to overlap with the ideal state.

Through simulation we also explored the limits of state recovery through purification and found that in low noise regimes, the noise floor correlates with the magic in the state. The upper bound of entanglement and magic of an $n$-qubit state each scale with $n$, and the general upper bound on the noise floor exponentially decreases with the Renyi entropy of the error probabilities~\cite{koczor2021dominant}. Future work would be to understand these bounds from the lens of the magic and entanglement of specific states of interest, and with well characterized hardware error. 

In an additional study, we demonstrate that a unitary with higher overlap to the target state does not necessarily lead to better performance if during its execution the magic it generates is not protected. We observed a decrease in performance when magic and entanglement were generated early in the construction of the state, resulting in a prolonged exposure to noise. A natural extension of our study would be to design protocols for an adaptive ansatz such as iterative qubit coupled cluster~\cite{ryabinkin2020iterative} or ADAPT-VQE~\cite{grimsley2019adaptive} to choose successive operators which maximize late generate of resources as opposed to those with highest energy gradient. Since the noise floor is difficult to determine apriori, one could potentially mitigate coherent mismatch by considering an algorithm which randomizes over a set of possible paths in a similar way to randomized compiling~\cite{wallman2016noise}.

The re-use of experimental data was shown by tailoring an entire $H_2$ dissociation curve to be computable by a single circuit. The best performing could be identified through monitoring errors rate, and also corresponded to the state with the lowest errors in magic. Our findings support future work on classical or quantum methods to extract information and utility from noisy quantum states, and to use quantum informatic tools to influence quantum algorithm development. 

\section{Methods}
\subsection{Simulations and Hamiltonian preprocessing details}

\subsubsection{Tailoring energy calculations to lowest error circuit}
\label{sec:tailoring_hamiltnoian}

If an application requires calculating Pauli expectation values, there is flexibility in how that information is extracted~\cite{gunlycke2024cascaded}. For example, consider a qubit Hamiltonian $H=\sum_ic_iP_i$ representing the electronic structure of chemical system, and its groundstate is generated by the circuit $|\Psi\rangle = U_{ansatz}|0\rangle$. The electronic ground state energy can be estimated as: 
\begin{equation}
 E = \langle 0 | U^\dag_{ansatz} H U_{ansatz}| 0 \rangle.
\end{equation}
To calculate  E with a circuit $U_{circ}$ that does not correspond to $U_{ansatz}|0\rangle$, we can find an appropriate dressing of the Hamiltonian $H^*=D^\dag H D$ such that 
\begin{equation}
 E = \langle0|U^\dag_{ansatz} H U_{ansatz}|0\rangle =\langle0|U^\dag_{circ} D^\dag H DU_{circ}|0\rangle = \langle 0 | U^\dag_{circ} H^* U_{circ}| 0 \rangle.
\end{equation}
Identifying the optimal operator $D$ introduces additional complexity, but we will consider the scenario where both $U_{ansatz}$ and $U_{circ}$ are constructed using Pauli exponential operators. For example, in the variational framework, for a given circuit ansatz $U_{ansatz}(\theta)$, the variational parameters can be optimized to the set $\theta^*$ to calculate the ground state energy of a Hamiltonian. 
For circuits constructed using products of Pauli exponentials, $D$ can be written as:
\begin{equation}
 D=\prod_{m}e^{-i\theta^*_m P_m}\prod_{\mu}e^{i\delta_\mu P_\mu}.
\end{equation}
Where $\theta^*_m$ are the original parameters from the optimized ansatz operators $P_m$, and $\delta_{\mu}$ are the parameters of the chosen circuit operators $P_{\mu}$.

Although dressing of the Hamiltonian increases the number of terms, this can be mitigated by choosing points where the unitaries become Clifford \cite{beguvsic2023simulating} or by writing the exponentials as first order taylor expansions as is done in \cite{liu2022reducing}. We also note that if you are operating with a fixed circuit, then any expectation value can be reused for a different calculation. In that case, even if the number of Hamiltonian terms for a single system may grow, if you were to calculate many systems, then it may results in far fewer measurements.

\subsubsection{Method of increments with frozen natural orbitals (MI-FNO)}
\label{sec: MIFNO}
The method of increments (MI)~\cite{nesbet1967atomic} tackles a many-body problem by considering it as a summation of a collection of smaller n-body problems (generally up to 3-body problems are considered). Each of these problems will be referred to as fragment in the following text, and they are computationally cheaper to treat via quantum chemistry.

Concretely, the Be atom was expanded in terms of occupied orbitals, and MI have employed to systematically reduce the occupied orbital space. To address the virtual orbital space truncation, we use the frozen natural orbitals (FNO) method as described in~\cite{verma2021scaling}. This process has been applied prior to the quantum chemistry calculation for every fragment. It consists on computing the one-particle density matrix from a cheap second-order many-body perturbation theory (MP2), and then neglecting a subset of the virtual orbital space below an electron-occupancy threshold (a correction term can be added to improve the accuracy, thanks to the MP2 calculation). The resulting fragments are thus a combination of different active spaces from the full molecular problem. The final energy is computed from the summation of all the n-body energy increments, as depicted in~\cite{verma2021scaling}. For the Be atom, we took the virtual space to be 2, and found that only one fragment carried electron correlation. This fragment was used in the experiments, and we report only the fragment energy, not the total energy. 

\subsubsection{$H_3$ unitary approximations}
\label{sec: unitary_paths_details}
\begin{figure*}[hbt!]   \includegraphics[scale=0.65]{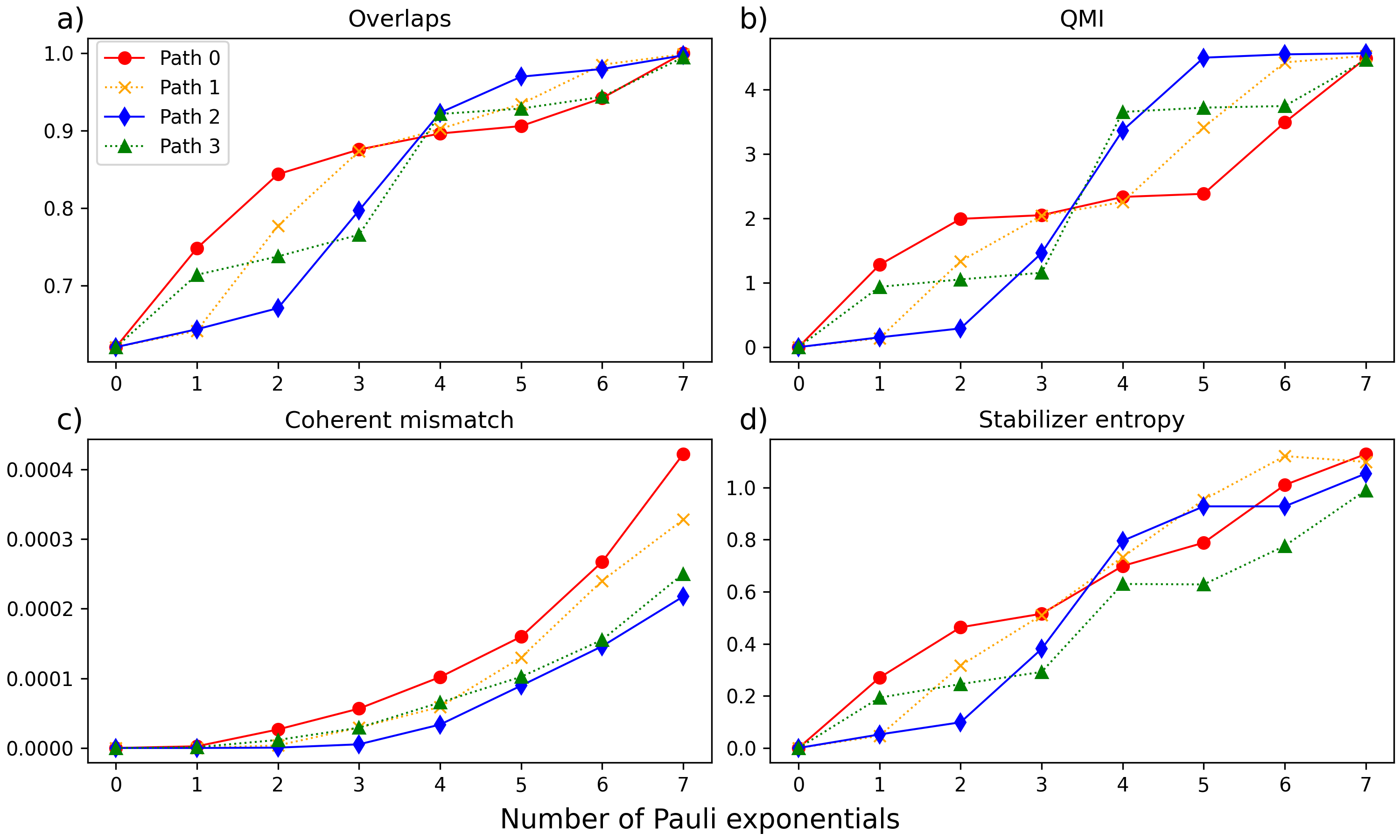}
    \caption{Metrics for four unitary approximations to the ground state of a linear $H_3$ molecule after the application each Pauli exponential $g_i$, where number of exponentials is marked on the x-axis. a) Overlap with the target state. b) The quantum mutual information (QMI). c) Coherent mismatch. d) The stabilizer entropy  ($W_2$) }
    \label{fig:circuit_path_all}
\end{figure*}

\begin{figure*}[hbt!]   \includegraphics[scale=0.65]{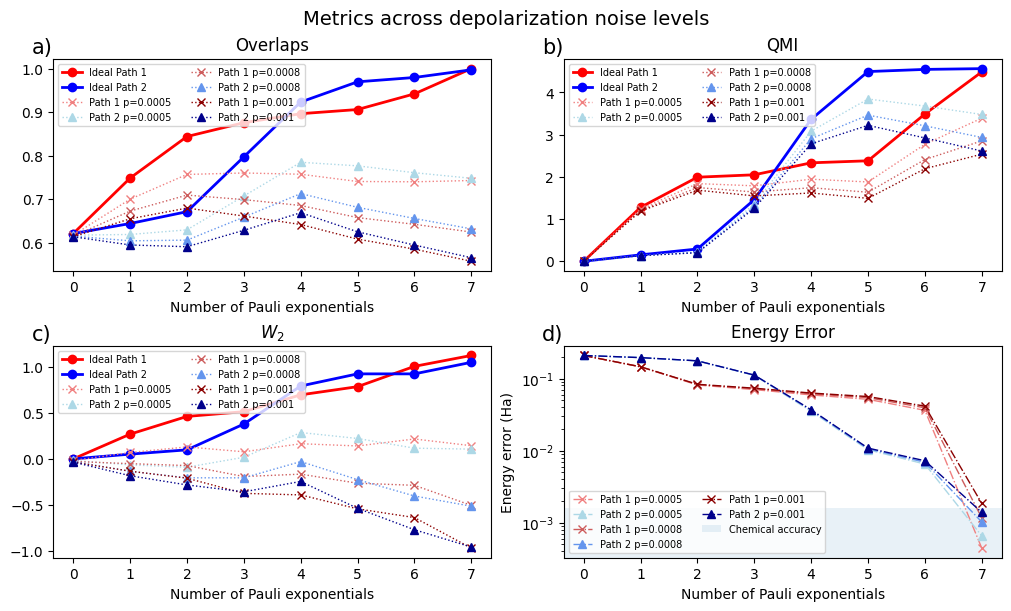}
    \caption{Metrics for two paths (red and blue) which generate approximations to the ground state of a linear $H_3$ molecule.  Metrics are measured after the application of each Pauli exponential $g_i$ where $n$ is marked on the x-axis. In figure a)-c), the solid lines represent the noiseless simulation, and the dotted lines are simulations with a range depolarization noise strengths from 0.0005 to 0.001 applied after each Pauli exponential. a) Overlap with the target state. b) The quantum mutual information (QMI). c) The Magic witness ($W_2$) d) Coherent mismatch and the energy error of the purified state. The grey band represents values within a chemical accuracy of $1.59$ mHa. }
    \label{fig:circuit_paths_depol}
\end{figure*}

\begin{figure*}[hbt!]   \includegraphics[scale=0.65]{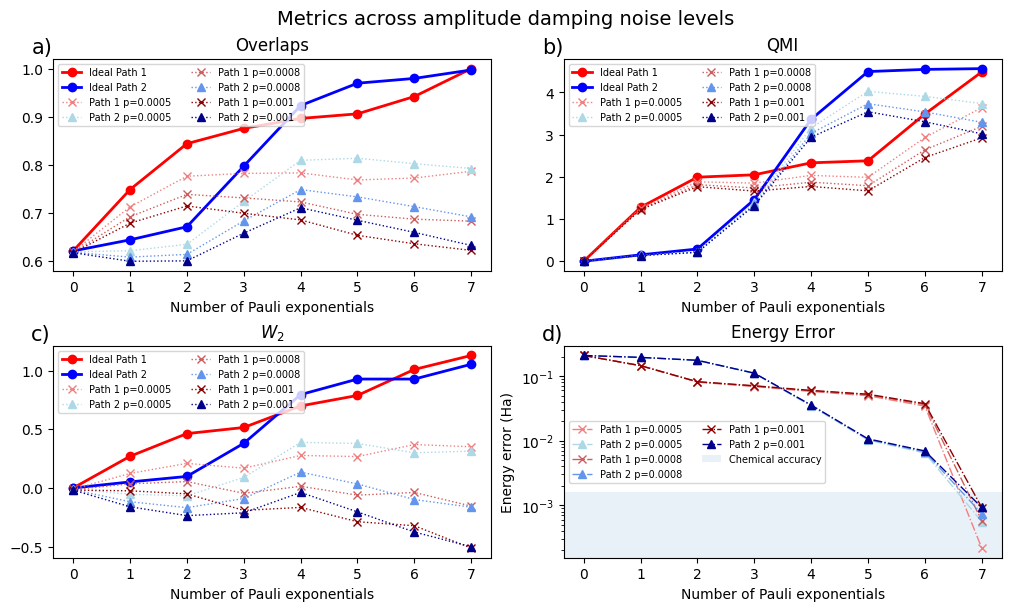}
    \caption{Metrics for two paths (red and blue) which generate approximations to the ground state of a linear $H_3$ molecule.  Metrics are measured after the application of each Pauli exponential $g_i$ where $n$ is marked on the x-axis. In figure a)-c), the solid lines represent the noiseless simulation, and the dotted lines are simulations with an amplitude damping noise strengths from 0.0005 to 0.001. a) Overlap with the target state. b) The quantum mutual information (QMI). c) The Magic witness ($W_2$) d) Coherent mismatch and the energy error of the purified state. The grey band represents values within a chemical accuracy of $1.59$ mHa. }
    \label{fig:circuit_paths_amp}
\end{figure*}

The linear $H_3$ molecule used an STO-3G basis, with a 2 Angstrom separation between each atom. We used the symmetry preserving Bravyi Kiteav~\cite{bravyi2017tapering, seeley2012bravyi} fermion to qubit encoding which resulted in a 6 qubit system. In Table~\ref{tab:pauli_exponentials} we list four different unitary approximations to the target state. In Fig~\ref{fig:circuit_path_all} we show the relevant metrics for each resulting circuit path.  We also include more noisy simulations with depolarization noise in~Fig.~\ref{fig:circuit_paths_depol} and amplitude damping noise channels in Fig.~\ref{fig:circuit_paths_amp}.

\begin{table}[ht]
\centering
\footnotesize
\begin{tabular}{c >{\centering\arraybackslash}p{10cm}}
\toprule
\textbf{Path} & \textbf{Pauli Exponentials} \\
\midrule
0 \\[0.5ex]
Circuit ordering for $U_1$
& 
\begin{tabular}[t]{@{}l@{}}
$0.243\,[Y_0\, Z_1\, X_2\, X_3\, X_5],\; 0.197\,[Y_1\, X_2\, X_3\, X_4],\; -0.0965\,[Y_0\, X_2\, Z_3]$ \\[0.5ex]
$0.183\,[Y_3\, X_5],\; -0.0965\,[Y_0\, X_2\, Z_3],\; 0.197\,[Y_1\, X_2\, X_3\, X_4]$ \\[0.5ex]
$0.243\,[Y_0\, Z_1\, X_2\, X_3\, X_5]$
\end{tabular} \\
\midrule
1 & 
\begin{tabular}[t]{@{}l@{}}
$0.0913\,[Y_3\, X_5],\; 0.243\,[Y_0\, Z_1\, X_2\, X_3\, X_5],\; 0.197\,[Y_1\, X_2\, X_3\, X_4]$ \\[0.5ex]
$-0.193\,[Y_0\, X_2\, Z_3],\; 0.197\,[Y_1\, X_2\, X_3\, X_4],\; 0.243\,[Y_0\, Z_1\, X_2\, X_3\, X_5]$ \\[0.5ex]
$0.0913\,[Y_3\, X_5]$
\end{tabular} \\
\midrule
2\\[0.5ex]
Circuit ordering for $U_2$& 
\begin{tabular}[t]{@{}l@{}}
$-0.0965\,[Y_0\, X_2\, Z_3],\; 0.0913\,[Y_3\, X_5],\; 0.243\,[Y_0\, Z_1\, X_2\, X_3\, X_5]$ \\[0.5ex]
$0.394\,[Y_1\, X_2\, X_3\, X_4],\; 0.243\,[Y_0\, Z_1\, X_2\, X_3\, X_5],\; 0.0913\,[Y_3\, X_5]$ \\[0.5ex]
$-0.0965\,[Y_0\, X_2\, Z_3]$
\end{tabular} \\
\midrule
3 & 
\begin{tabular}[t]{@{}l@{}}
$0.197\,[Y_1\, X_2\, X_3\, X_4],\; -0.0965\,[Y_0\, X_2\, Z_3],\; 0.0913\,[Y_3\, X_5]$ \\[0.5ex]
$0.486\,[Y_0\, Z_1\, X_2\, X_3\, X_5],\; 0.0913\,[Y_3\, X_5],\; -0.0965\,[Y_0\, X_2\, Z_3]$ \\[0.5ex]
$0.197\,[Y_1\, X_2\, X_3\, X_4]$
\end{tabular} \\
\bottomrule
\end{tabular}
\caption{Pauli exponentials and $\theta$ values for different unitary approximations.}
\label{tab:pauli_exponentials}
\end{table}

\subsubsection{Purification order}
\begin{figure*}[hbt!]   \includegraphics[scale=0.8]{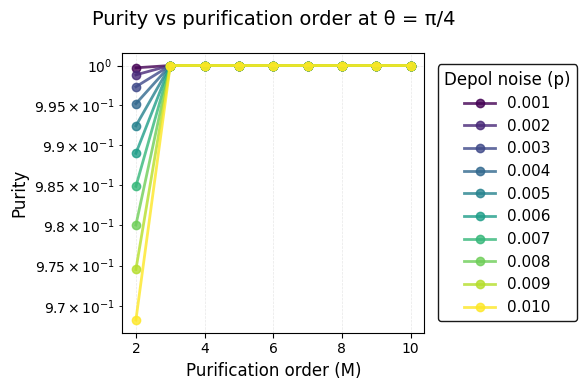}
    \caption{Purity of state vs order of purification for a range of depolarization noise strengths for the $|\psi\rangle=e^{-i\theta YXXX}|1100\rangle$ state at $\theta = \pi/4$. }
    \label{fig:purification_order}
\end{figure*}

In Fig~\ref{fig:purification_order} we show the purity of the state $|\psi\rangle=e^{-i\frac{\pi}{4} YXXX}|1100\rangle$ after purification for M up to 10 and for a range of depolarization noise strengths. The purity converges around $M=3$, and we take $M=5$ for this study.  

\subsubsection{Simulation backends and packages}
All simulations used cirq~\cite{CirqDevelopers_2024} density matrix simulator and QutTiP~\cite{johansson2012qutip} was used to calculate overlaps and Von Neumann entropy. Tangelo~\cite{senicourt2022tangelo} was used to define molecular systems, encoding them into qubits, and generate circuits.   

\subsection{Experimental details}
\label{sec: exp_details}

A symmetry projection circuit, shown in Figure \ref{cct: postselect}, enables postselection by using an ancilla qubit to identify the parity of the state \cite{bonet2018low}, and removing data with the wrong parity. For all experimental data we approximate the density matrix $\rho$ using a single qubit Pauli classical shadow~\cite{huang2020predicting} constructed from the postselected data~\cite{jnane2024quantum}. This density matrix is used for purification~\cite{seif2023shadow} and to calculate the reported quantum resource metrics. For error bars on the ratio of postselected data, we calculated the average and standard deviation of ratio of postselected data for each basis set in each experiment (81 total).

\begin{figure}[h!]
\begin{quantikz}  
&\gate[4]{U(\theta)} &\qw & \qw &\qw  &\qw &\ctrl{4}& \qw \\
&&\qw & \qw & \qw & \ctrl{3} & \qw& \qw\\
&&\qw & \qw & \ctrl{2} & \qw & \qw& \qw \\ 
&&\qw & \ctrl{1} & \qw & \qw & \qw& \qw \\
&&\lstick{\ket{0}} & \targ{} & \targ{} &\targ{} &\targ{} &\meter{}
\end{quantikz}
\caption{Symmetry projection circuit used for postselection.}
\label{cct: postselect}
\end{figure}

\subsubsection{Randomized classical shadows}
The classical shadows method~\cite{huang2020predicting} was used in both the Aria noisy device simulation and the Aria experiment to gather data from the quantum computer. Specifically, we utilized the randomized classical shadows method, where the measurement bases were chosen randomly from all possible $3^{n_{\text{qubits}}}$ unitaries constructed from tensor products of random single-qubit Clifford circuits. 

The $2^{n_{\text{qubits}}} \times 2^{n_{\text{qubits}}}$ state was estimated by inverting the quantum channel corresponding to random single-qubit Clifford circuits for each measurement, as described by the following equation:
\begin{equation}
\hat{\rho} = \bigotimes_{j=1}^{n_{\text{qubits}}} \left( 3U_j^{\dagger} |b_j\rangle \langle b_j| U_j - \mathbb{I} \right)
\label{eq:cs_rho_estimate}
\end{equation}
where $U_j$ and $|b_j\rangle$ denote the single-qubit basis and measurement outcome, respectively. To obtain the estimated density matrix for the system, we averaged a set of $M$ snapshots $\{\hat{\rho}_1, \hat{\rho}_2, ..., \hat{\rho}_M\}$. Arbitrary observables were computed from the median of means prediction, as described in Ref.~\citenum{huang2020predicting}. According to quantum information theory, the error $\epsilon$ expected from an observable estimation requires at least $\log{(M)}\ \max_i\|O_i\|^2_{\text{shadow}} / \epsilon^2$ measurements, where $M$ is the number of linear functions and $\|O_i\|^2_{\text{shadow}}$ refers to the locality of $O_i$.

Aria noisy device simulation and all experimental data presented were on 4 qubit systems, where we took 1000 shots for each Pauli basis, and experimental data was additionally postselected. To calculate error bars, we bootstrapped the classical shadows 250 times. The reported metrics are the mean values of calculated by the 250 shadows, and the error bars are the standard deviation. 

\subsubsection{Experimental circuits}
\label{sec: exp_circuit_details}

For the deeper circuit results in Fig.~\ref{fig:Be_depth} and ~\ref{fig:exp_path_SE_QMI}, when the number of Pauli exponentials $N$ is 1, the experiment was a two qubit circuit defined by $e^{-i0.0.401YX}|11\rangle$, and the other two qubits were simulated classically. When $N=2$, the experiment run was the circuit corresponding to $e^{-i0.334YZXZ}e^{-i0.334YXXX}|1111\rangle$, and when $N=3$, the circuit run was defined by  $e^{-i0.0.200YZXZ}e^{-i0.400XYXX}e^{-i0.0.2007YZXZ}|1111\rangle$. The first two circuits were constructed using the qubit coupled cluster method~\cite{ryabinkin2020iterative}, and the third circuit used the ILC method~\cite{ryabinkin2023efficient}, both implemented in Tangelo~\cite{senicourt2022tangelo}.

\section{Acknowledgments}
This work was supported as part of a joint development agreement between Dow and SandboxAQ. We are grateful to Peter Margl from Dow for the many technical discussions, and for his feedback on the results. We also thank Joshua Goings at IonQ for his assistance in running the experiments on Aria, for insightful discussions on the results, and for his valuable feedback on the manuscript. 

\section{Data availability statement}
The data that support the experimental findings of this study are openly available at the following URL/DOI: \hyperlink{}{https://doi.org/10.5281/zenodo.19868803}
\bibliography{bib}

\end{document}